\documentclass[journal=ancac3, manuscript=article, layout=twocolumn, fleqn]{achemso}

\usepackage{chemformula} 
\usepackage[T1]{fontenc} 
\usepackage{stfloats} 
\usepackage{xargs}
\usepackage{amsmath,amssymb,amstext}
\usepackage{bbold} 
\usepackage{hyperref}
\usepackage{graphics}

\usepackage{xr}

\definecolor{C0}{rgb}{0.12156862745098,0.466666666666667,0.705882352941177}
\definecolor{C1}{rgb}{1,0.498039215686275,0.0549019607843137}
\definecolor{C2}{rgb}{0.172549019607843,0.627450980392157,0.172549019607843}
\definecolor{C3}{rgb}{0.83921568627451,0.152941176470588,0.156862745098039}
\definecolor{C4}{rgb}{0.580392156862745,0.403921568627451,0.741176470588235}
\definecolor{C5}{rgb}{0.549019607843137,0.337254901960784,0.294117647058824}
\definecolor{C6}{rgb}{0.890196078431372,0.466666666666667,0.76078431372549}
\colorlet{C7}{gray!99.607843137254903!black}
\definecolor{C8}{rgb}{0.737254901960784,0.741176470588235,0.133333333333333}
\definecolor{C9}{rgb}{0.0901960784313725,0.745098039215686,0.811764705882353}

\usepackage[colorinlistoftodos, prependcaption, textsize=tiny]{todonotes}
\newcommandx{\LH}[2][1=]{\todo[linecolor=C3,backgroundcolor=C3!25,bordercolor=C3,#1]{\color{C3}\bf LH: #2}}
\newcommandx{\AJ}[2][1=]{\todo[linecolor=C2,backgroundcolor=C2!25,bordercolor=C2,#1]{\color{C2}\bf AJ: #2}}
\newcommandx{\OTH}[2][1=]{\todo[linecolor=C0,backgroundcolor=C0!25,bordercolor=C0,#1]{\color{C0}\bf OTH: #2}}

\newcommand{\angstrom}{\mbox{\normalfont\AA}}

\emergencystretch=\maxdimen
\newcommand{\onstate}{on-state}
\newcommand{\offstate}{off-state}
\newcommand{\mixedstate}{mixed-state}

\newcommand{\Onstate}{On-state}
\newcommand{\Offstate}{Off-state}
\newcommand{\Mixedstate}{Mixed-state}

\newcommand{\onoffstate}{on- and off-state}
\newcommand{\Onoffstate}{On- and off-state}

\newcommand{\onmixedstate}{on- and mixed-state}

\newcommand{\offmixedstate}{off- and mixed-state}

\newcommand{\onoffmixedstate}{on-, off- and mixed-state}

\newcommand{\WF}{$\Delta \Phi$}
\newcommand{\WFs}{$\Delta \Phi$s}

\newcommand{\geom}{geometry}
\newcommand{\geoms}{geometries}

\newcommand{\structure}{motif}
\newcommand{\structures}{motifs}
\newcommand{\Structures}{Motifs}

\newcommand{\SAMPLEGPR}{SAMPLE-GPR}

\author{Lukas H\"ormann}
\author{Andreas Jeindl}
\author{Oliver T. Hofmann}
\email{o.hofmann@tugraz.at}
\affiliation[Graz University of Technology]
{Institute of Solid State Physics, Graz University of Technology, Petersgasse 16, 8010 Graz, Austria}

\title[]
	{From a bistable adsorbate to a switchable interface: tetrachloropyrazine on Pt(111)}

\begin{document}

\begin{tocentry}
\includegraphics[width=1\linewidth]{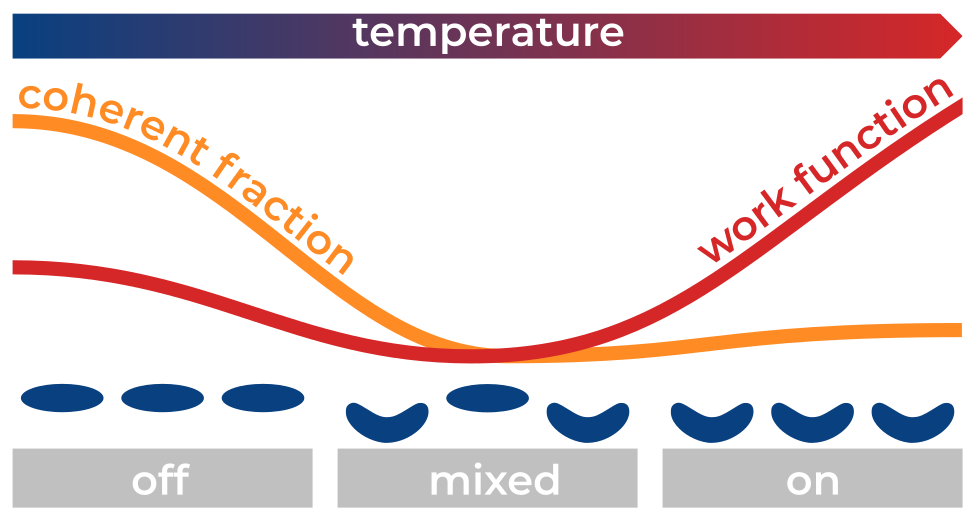}
\end{tocentry}

\twocolumn[
\begin{@twocolumnfalse}
\begin{abstract}
Virtually all organic (opto)electronic devices rely on organic/inorganic interfaces with specific properties. These properties are, in turn, inextricably linked to the interface structure. Therefore, a change in structure can introduce a shift in function. If this change is reversible, it would allow constructing a switchable interface. We accomplish this with tetrachloropyrazine on Pt(111), which exhibits a double-well potential with a chemisorbed and a physisorbed minimum. These minima have significantly different adsorption geometries allowing the formation of switchable interface structures. Importantly, these structures facilitate different work function changes and coherent fractions (X-ray standing wave measurements), which are ideal properties to readout the interface state.

We perform surface structure search using a modified version of the SAMPLE approach and account for thermodynamic conditions using ab-initio thermodynamics. This allows investigating millions of commensurate as well as higher-order commensurate interface structures. We identify three different classes of structures exhibiting different work function changes and coherent fractions. Using temperature and pressure as handles we demonstrate the possibility of reversible switching between those different classes, creating a dynamic interface for potential applications in organic electronics.
\end{abstract}
\end{@twocolumnfalse}
]


\section{Introduction}
Organic/inorganic interfaces are essential for the construction of organic (opto)electronic devices. To date research has mainly focused on attaining specific interface properties through controlling the structure and chemistry of the organic adlayer.\cite{casalini2017self, slowinski1997through, rivnay2018organic, katz2009thin, wu2018strategies} In this theoretical study we go beyond these efforts and study organic adsobate layers with switchable properties.

Switchable interfaces in literature rely on several function principles including modifying interactions at the interface\cite{li2019deep} and applying external stimuli such as optical signals\cite{blinov2002photoreactive, szymanski2013azobenzene}, electric fields\cite{fujii2019electric, tam2010reversible}, magnetic fields\cite{willner2003magnetic}, temperature,\cite{song2010off} biochemical processes\cite{tam2008biochemically} or pH-value\cite{bocharova2012switchable}. Here, we use temperature and pressure to switch adlayers of molecules that exhibit a double well potential when adsorbing on a substrate. Examples for systems with double well potentials include benzene derivatives on Pt(111) and anthradithiophene on Cu(111).\cite{LiuWei_bistable_molecules, borca2015bipolar, schendel2016remotely} These systems show great potential as molecular switches\cite{LiuWei_bistable_molecules} and have even been suggested as data storage.\cite{chen2020conductance} Here, we focus on the system of tetracholoropyrazine (TCP) on Pt(111), which displays a double-well potential on the Pt(111) surface.\cite{LiuWei_bistable_molecules} The two types of adsorption geometries, called {\onoffstate}, exhibit different types of bonding and adsorption geometries. Crucially, both minima have approximately the same adsorption energy, allowing the formation of diverse surface structures which are sufficiently close in energy to be reversibly switched with external stimuli. Interestingly, in this work we find that individual chemisorbed molecules are energetically more favorable than physisorbed ones. However, this ordering reverses in continuous layers (at low temperatures).

Reading out the interface states requires that the switch in structure entails a shift in interface properties. We focus on two of these properties, namely the work-function change {\WF} (compared to the clean surface) and the coherent fraction (as obtained by X-ray standing wave methods), and study how to modify them using environmental stimuli such as temperature and pressure. Different possible polymorphs exhibits diverse {\WFs} and coherent fractions. However, the amount by which they can be switched is significantly smaller the total range. We identify the reasons for this and provide ideas to overcome it.

Molecules on surfaces may arrange into a large number of possible polymorphs. Therefore, we investigate the {\WFs} and the coherent fractions of millions of surface structures. To determine these structures and their adsorption energies we use the SAMPLE approach,\cite{sample} which combines dispersion-corrected density functional theory (DFT) with machine learning (see Methods Section). Further, ab-initio thermodynamics allows modeling the impact of temperature and pressure on the surface polymorphism. This enables us to conduct in-silico simulations of experiments.

A number of experimental methods exist to determine work functions (and in turn {\WFs}), such as photo-emission spectroscopy and the Kelvin probe method. The change in work function resulting from molecules adsorbing on a surface depends on a number of factors. These include pushback of electron density into the surface, charge transfer across the interface, formation of new interface states and bonds between adsorbates and the substrate, image charge effects, permanent dipoles of the molecules as well as the coverage.\cite{zojer2019impact, OTERO2017105, braun2009energy} {\Onoffstate} differ significantly in type of bonding and adsorption geometry. Therefore, we expect that adsorbate layers consisting of {\onoffstate} adsorption geometries yield dissimilar {\WFs}, allowing the construction of a switchable interface.

A useful experimental method to determine the adsorption geometry is the X-ray standing wave (XSW) technique.\cite{xsw_1987} This method yields two measures, namely the coherent position and the coherent fraction. The coherent position allows determining the mean adsorption height of the adsorbates. The coherent fraction is effectively an order parameter\cite{xsw_2020} containing information about differences in adsorption heights, which would occur if {\onoffstate} molecules coexisted in a particular interface state. If all atoms (of a particular species) within all molecules adsorb at an identical adsorption height, the coherent fraction is $1$, while it decreases with variations in adsorption height. Thus, the coherent fraction will allow differentiating between {\onoffmixedstate} layers.

The fact that both {\WF} and the coherent fraction are experimentally readily accessible makes these properties ideal candidates to readout the state of a switchable interface. 


\section{Results}

\subsection{Individual Molecules on the Surface}

When investigating the adsorption of single molecules, we are interested in local minima of the potential energy surface (PES) which constitute energetically favorable adsorption geometries. We will hereafter refer to these as ``{\geoms}''. As stated above, TCP on Pt(111) exhibits a double well potential, with minima occurring at two different adsorption heights. Hence, two different types of adsorption geometries exist, namely {\onoffstate} {\geoms}. While {\onstate} {\geoms} are chemisorbed, {\offstate} {\geoms} are physisorbed.

We determine the {\geoms} in two steps: Initially, a Gaussian-process regression (GPR) algorithm identifies the minima geometries of a coarse grained PES (see Methods Section). The algorithm uses DFT-calculated energies of a few adsorption geometries as input and interpolates between them. Hereby, it only considers the most important degrees of freedom, i.e., position, orientation and bending of the molecule (see Methods Section). If necessary (see below), we refine the GPR minima using DFT geometry optimizations.

\begin{figure}[tbph!]
	\includegraphics[width=1\linewidth]{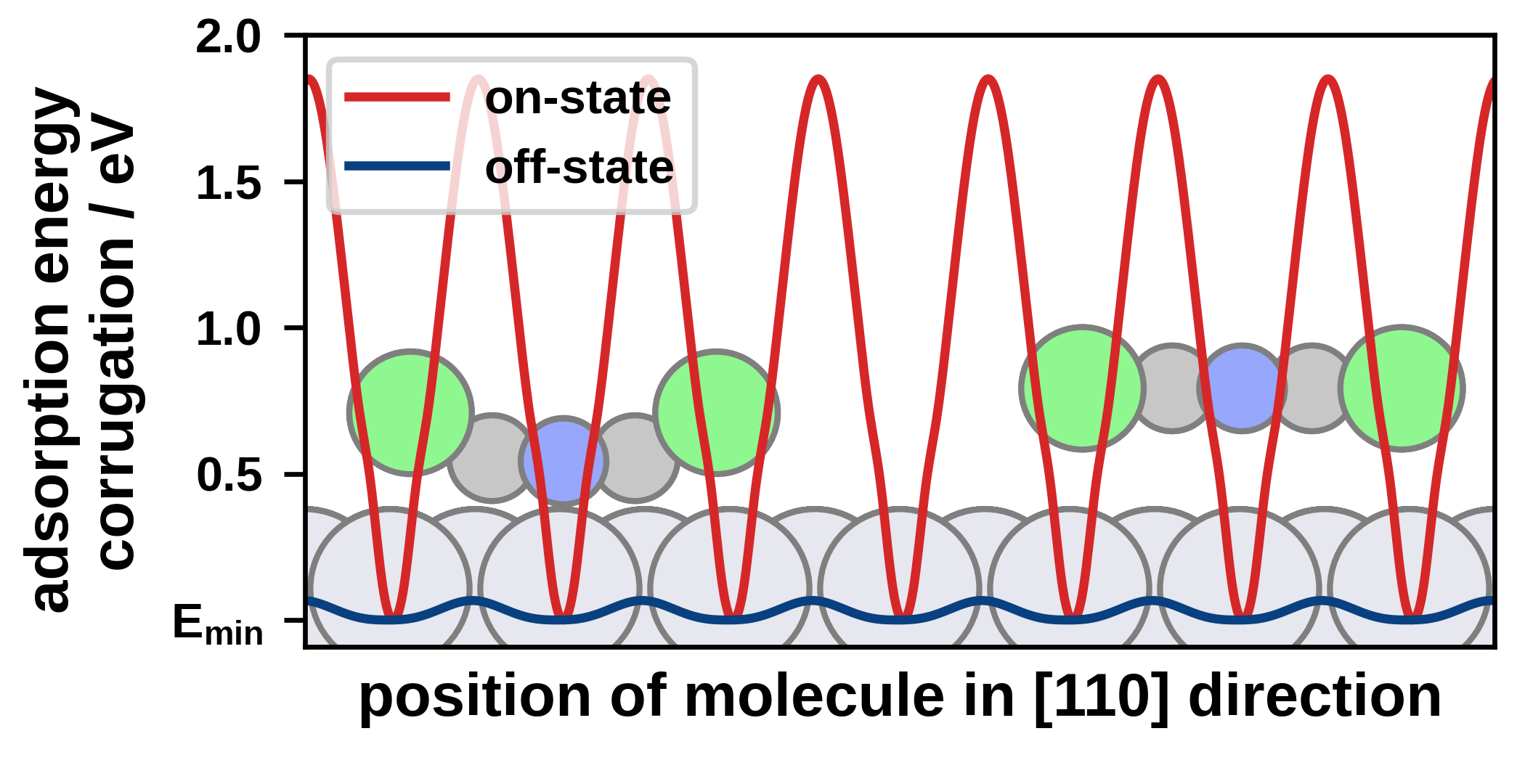}
	\caption{Cut through the GPR PES of TCP on Pt(111) in [110] direction; (red) PES of the {\onstate} {\geom}; (blue) PES of the {\offstate} {\geom}; energies are plotted relative to the respective minimum energy $E_{min}$.}
	\label{fig:pes1d}
\end{figure}

For {\offstate} {\geoms} the coarse grained PES is sufficiently accurate since the molecules mainly bond via spatially uniform van der Waals interactions. This leads to a weakly corrugated PES, with an energy range of approximately $0.2~eV$ (see Figure \ref{fig:pes1d}). Hence, we can directly use the GPR minima as adsorption geometries. In the {\offstate} the molecules remains flat and has five adsorption geometries with (GPR-calculated) bonding energies of $-0.97~eV$ to $-0.95~eV$ and adsorption heights of approximately $3.3~\angstrom$. DFT geometry optimizations of these minima yield only very small changes in geometry (molecule remains flat) and the gain in adsorption energy is within the uncertainty of our method.

Conversely, the PES of {\onstate} {\geoms} is strongly corrugated, with minima and maxima spanning an energy range of approximately $2.0~eV$ (see Figure \ref{fig:pes1d}). This is due to a pronounced spatial dependence of the covalent interactions between the molecule and the surface. The molecules are strongly bent, with the Cl atoms pointing away from the surface. Due to the higher complexity of the {\onstate}, a PES considering only position, orientation and bending of the molecule is insufficient to accurately determine adsorption geometries. Hence, we refine the GPR minima with DFT, allowing the molecule and the substrate atoms of the two topmost layers to relax. Hereby, the relaxation of substrate atoms below the molecule contributes up to $-0.5~eV$ to the adsorption energies (more negative is more bonding). For the {\onstate} we find four different adsorption geometries that have adsorption energies (calculated with DFT) of $-1.07~eV$ to $-1.04~eV$ and an adsorption height of approximately $2.1~\angstrom$ above the unrelaxed substrate (This is how an XSW-experiment would determine the adsorption height.\cite{xsw_1987}).

For visualizations of the potential energy surfaces and additional details regarding adsorption geometries see the Supporting Information.

\subsection{Interface Structures}

Having discussed the adsorption geometries of individual molecule on the surface, we will now focus on the close-packed layers they form. In general, adlayers assume different types of commensurability depending on a delicate balance between molecule-molecule interactions and the corrugation of molecule-substrate interactions.\cite{hooks2001epitaxy, mannsfeld2005understanding} Strong molecule-molecule interactions and a comparatively weak corrugation of molecule-substrate interactions allow maximizing the energy gain from interactions between molecules. This will most likely lead to incommensurate layers. Conversely, a large corrugation of molecule-substrate interactions and comparatively weak molecule-molecule interactions forces the molecules to remain in energetically favorable adsorption sites. Concurrently, any energy gain from favorable molecule-molecule interactions would be outweighed by the energy penalty from unfavorable molecule-substrate interactions. This case leads to commensurate layers. Using the arguments from the previous paragraph, we can anticipate if an adlayer will be commensurate or incommensurate.

The two types of local adsorption geometries can form three different classes of adlayer structures. We will hereafter refer to a close-packed adlayer of molecules as a ``{\structure}''.

First, {\offstate} {\structures} consist purely of {\offstate} {\geoms}. Here, the PES of the single molecule is weakly corrugated (see Figure \ref{fig:pes1d}). Hence, the molecule can seek out the most beneficial molecule-substrate interactions, making {\offstate} {\structures} incommensurate.

Next, {\onstate} {\structures} consist only of {\onstate} {\geoms}. The PES of {\onstate} {\geoms} is strongly corrugated (see Figure \ref{fig:pes1d}). Therefore, the molecules seek the most favorable molecule-substrate interaction, which leads {\onstate} {\structures} to be commensurate.

Finally, {\mixedstate} {\structures} contain both {\onoffstate} {\geoms}. Here, the {\offstate} {\geoms} seek out the most beneficial molecule-substrate interaction, while the {\onstate} {\geoms} remain stuck due to the strongly corrugated molecule-substrate interactions. Hence, we expect {\mixedstate} {\structures} to be commensurate.

To find the energetically most favorable {\structures}, we first perform commensurate structure search using the SAMPLE approach (see Methods Section).\cite{sample} Hereby, we generate all possible commensurate {\structures} with different coverages and up to three molecules per unit cell and predict their adsorption energies. However, as we state above, we expect that some energetically favorable adlayers are incommensurate. Truly incommensurate {\structures} contain an infinite number of molecules per unit cell, each with a different adsorption site. This makes it obviously impossible to determine their energies from first principles. Nevertheless, we can approximate incommensurate {\structures} using higher-order commensurability. To consider higher-order commensurate {\structures}, we use a generalized SAMPLE approach ({\SAMPLEGPR}) based on a GPR model (see Methods Section). This approach also provides energies for geometry optimizations with limited degrees of freedom.

These additional capabilities are necessary to describe higher-order commensurate {\offstate} {\structures} and to optimize molecule positions in {\mixedstate} {\structures}, while the SAMPLE approach would be sufficient for {\onstate} {\structures}. For a consistent treatment of all three {\structure} classes, we rerank the $1000$ energetically most favorable {\structures} of every type and considered coverage (in total approximately $37000$ {\structures}) via {\SAMPLEGPR} and consider only {\SAMPLEGPR} energies hereafter.

For the {\offstate} we expect that incommensurate {\structures} are most energetically favorable. Hence, we use {\SAMPLEGPR} to optimize the $20$ most energetically favorable {\structures} using the unit cell parameters as well as location and orientation of the molecules as degrees of freedom (see Methods Section).

In {\mixedstate} structures the physisorbed molecules can move relatively freely. Therefore, we optimize their positions and orientations in the $40$ energetically most favorable {\mixedstate} {\structures} of every coverage, while keeping the chemisorbed molecules fixed.

\begin{figure}[!htbp]
	\centering
	\includegraphics[width=1\linewidth]{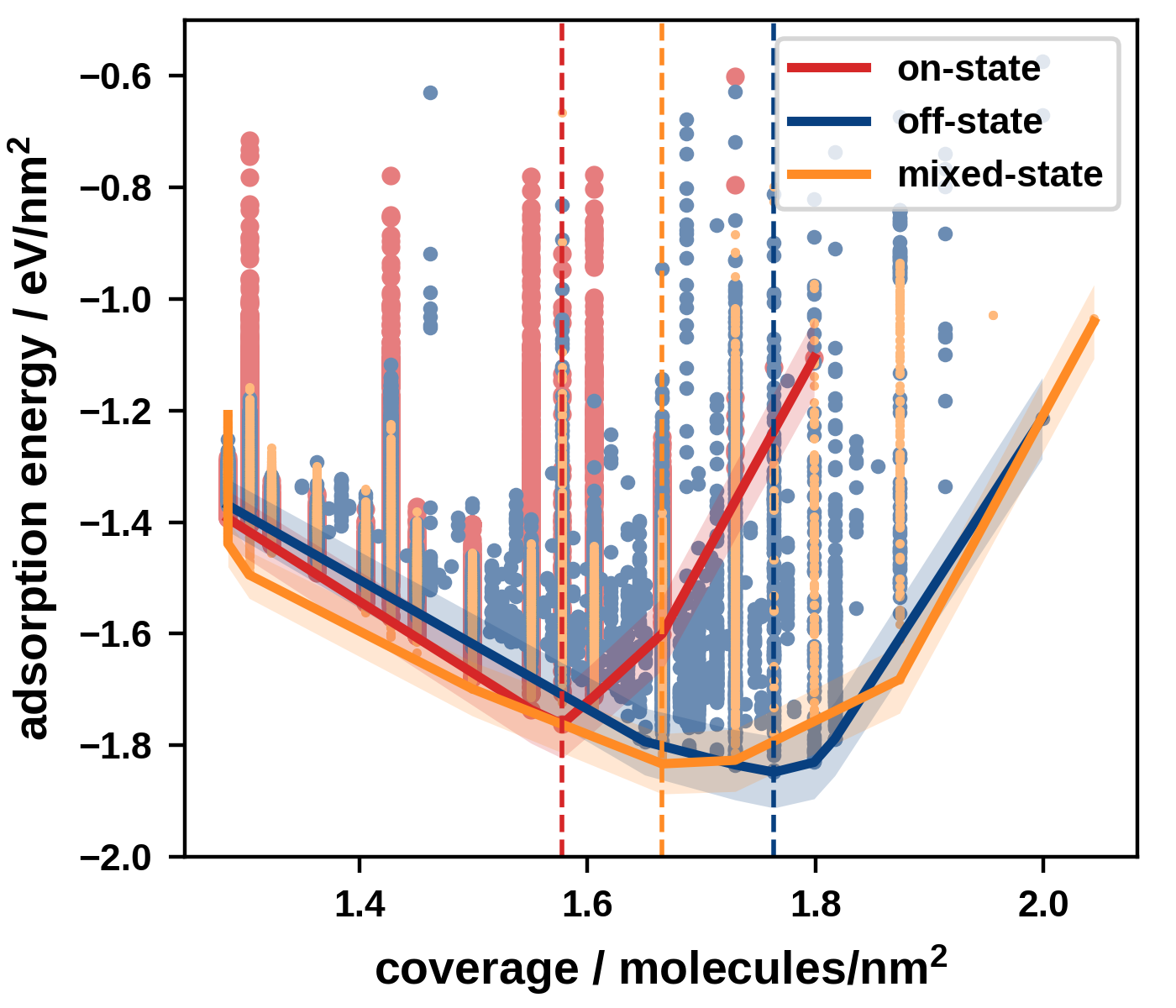}
	\caption{Adsorption energies per area for the $1000$ energetically most favorable, as well as the optimized {\onoffmixedstate} {\structures} at all considered coverages; the lines show the convex hull of the minimum energy; the shaded areas around the convex hulls show the prediction uncertainty.}
	\label{fig:energies_per_coverage}
\end{figure}

Figure \ref{fig:energies_per_coverage} shows these reranked (and optimized) energies per area plotted against the coverage for {\onoffmixedstate} {\structures}. In this chapter, we consider a temperature of $0~K$ where close-packed adsorbate layers seek to minimize the energy per area, making this the measure of interest.\cite{reuter2003first} At temperature above $0~K$, which we discuss in the next chapter, the Gibbs free energy of adsorption, which contains the energy per area as well as temperature-dependent potentials, becomes the relevant measure.\cite{rogal_reuter}

The different classes of {\structures} have their global minimum energy {\structure} at different coverages. For the {\offstate} the energetically most favorable {\structures} are higher-order commensurate, strongly indicating a preference of incommensurate layers. This allows {\offstate} {\structures} to pack more densely than {\onmixedstate} {\structures}. The energetically most favorable {\offstate} {\structure} (and overall energetically most favorable {\structure}) has a coverage of $1.76~molecules/nm^{2}$. {\Mixedstate} {\structures} also profit from closer packing. Although they are commensurate, the {\offstate} {\geoms} can adjust their positions relatively freely, allowing for more beneficial molecule-molecule interactions and tighter packing than in {\onstate} {\structures}. {\Mixedstate} {\structures} have a minimum at a coverage of $1.67~molecules/nm^{2}$. {\Onstate} {\structures} have the lowest packing density with the energetic minimum occurring at a coverage of $1.58~molecules/nm^{2}$. This is due to the strongly corrugated PES of {\onstate} {\geoms} which prevents them from leaving their potential well and thus makes it difficult to achieve favorable molecule-molecule interactions (see Figure \ref{fig:pes1d}).

Isolated {\onstate} {\geoms} have an approximately $0.10~eV$ more attractive molecule-substrate interaction than {\offstate} {\geoms}. However, in a tightly packed layer the molecules exhibit attractive molecule-molecule interactions. These interactions outweigh the energy penalty accrued from {\offstate} {\geoms}. This makes densely packed {\offmixedstate} {\structures} energetically more favorable than {\onstate} {\structures}.

\subsection{Switchable Interface States}
\label{ssc:switchabel_interface_states}

Having discussed possible interface {\structures}, we will now investigate how to use them for a switchable interface. This requires controllably shifting the work-function change ({\WF}) or the coherent fraction by a measurable, or better, a technically applicable margin. Hereby, the coherent fraction is technologically less relevant, but allows validating the switching behavior, since it is not directly related to {\WF}. Conversely, {\WF} is highly relevant for devices\cite{WOS:000489089600004, li2021interface, chen2020interfacial} and will therefore serve as focus of this discussion. For instance, a change in {\WF} of $200~meV$ would correspond to the typical threshold voltage of a germanium diode. Therefore, the {\structures} should provide an equally large or larger variety of {\WFs}. To test this, we will present an overview of the possible {\WFs} and coherent fractions.

In {\offstate} {\structures} the molecules are flat and assume very similar adsorption heights. The geometric uniformity of different possible polymorphs leads to similar {\WFs} with differences being largely due to coverage. {\Structures} with similar coverage exhibit {\WFs} that vary by only $100~meV$ and the largest {\WF} is $-690~meV$. Due to the similar adsorption geometries all {\offstate} {\structures} exhibit coherent fractions of C-atoms (as well as N- and Cl-atoms) of close to $1.00$.

In {\onstate} structures the molecules are strongly bent. Moreover, the different {\onstate} {\geoms} exhibit different adsorption heights, bending and tilting. These dissimilarities result in {\structures} with similar coverages exhibiting {\WFs} that differ by as much as $600~meV$ from each other. The maximum {\WF} is $-1020~meV$. Furthermore, the distortions of the adsorption geometries result in different z-positions of the C-atoms, for which we therefore find coherent fractions as low as $0.76$. We note in passing that the impact of the internal molecular geometry on the coherent fractions is known and has been used to elucidate adsorption geometries.\cite{xsw_2020, mercurio2013quantification}

{\Mixedstate} {\structures} contain flat lying {\offstate} {\geoms} as well as strongly distorted {\onstate} {\geoms}. Due to different adsorption geometries, {\mixedstate} {\structures} of similar coverage exhibit {\WFs} that differ by as much as $500~meV$ from each other. The largest {\WF} is $-930~meV$. Since molecules sit at different adsorption heights, the z-position of C-atoms in {\offstate} {\geoms} is approximately $1.2~\angstrom$ higher than that in {\onstate} {\geoms}, which is about half of the Pt lattice distance ($2.35~\angstrom$). Therefore, individual {\structures} exhibit coherent fractions for C-atoms that are as low as $0.01$. Coherent fractions smaller than $1$ are sometimes taken to indicate a disordered structure.\cite{okasinski2004x, bedzyk1995order} However, our {\structures} are highly ordered and commensurate. While thought experiments have shown the possibility of such a behavior,\cite{xsw_2020} this is, to our knowledge, the first report of such low coherent fractions for a system of lying molecules which could, in principle, be observed in an experiment. This effect is seen most strongly for C- and N-atoms and is less pronounced for Cl-atoms.

For additional plots and details regarding {\WF} and the coherent fraction see the Supporting Information.

\begin{figure*}[!htbp]
	\centering
	\includegraphics[width=1\linewidth]{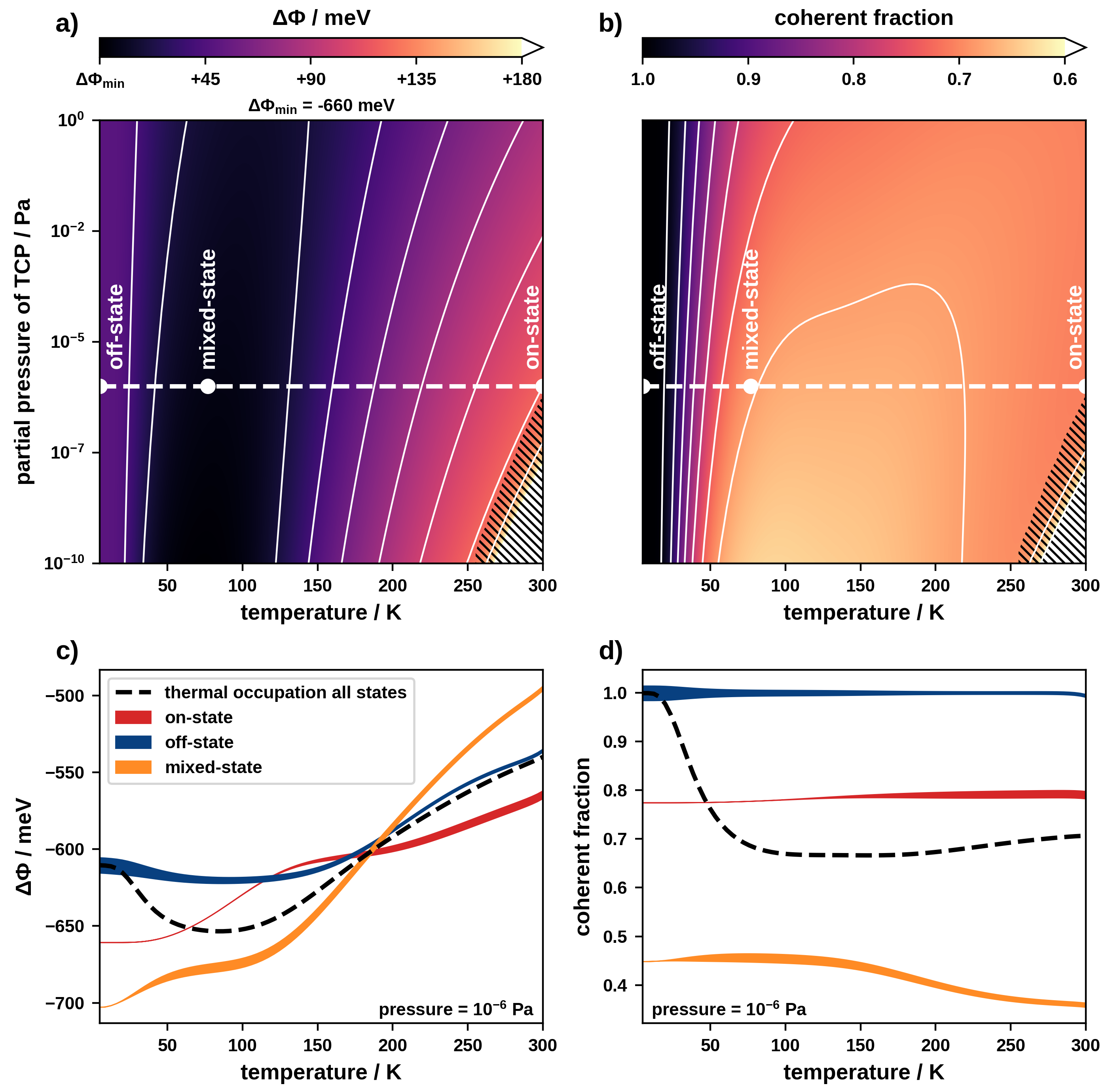}
	\caption{(a-b) Thermodynamically populated phase diagram showing (a) the expectation value of {\WF} and (b) the expectation value of the coherent fraction of C atoms, hatched areas indicate the thermodynamic range where adsorption is not energetically favorable; (c-d) cut through the phase diagram at a constant pressure of $10^{-6}~Pa$ showing expectation values for (c) {\WF} and (d) the coherent fraction separately for {\onoffmixedstate} {\structures}; the line width shows how much each type of {\structure} contributes to the thermal occupation of all states.}
	\label{fig:phase_diagram}
\end{figure*}

This discussion shows that {\onoffmixedstate} {\structures} theoretically have a large enough diversity in {\WFs} and coherent fractions to construct a switchable interface. Now we must find a way to control this diversity. An obvious strategy would be using temperature and pressure to shift the thermodynamic equilibrium and thereby influence which {\structures} form. In thermodynamic equilibrium the thermal occupation governs the probability of finding a particular {\structure}. In fact, multiple {\structures} may coexist on the surface and contribute to an average {\WF} or coherent fraction. Therefore, we must consider a set of {\structures}, rather than just the energetically most favorable one. We use a sufficiently large set containing the $37000$ energetically most favorable {\structures} as well as the clean Pt(111) surface (details in the Supporting Information). To include the influence of temperature and pressure in the adsorption energies of these {\structures} we use ab-initio thermodynamics.\cite{rogal_reuter} This allows generating the phase diagram which we show in Figure \ref{fig:phase_diagram}. Panels (a) and (b) show how the expectation values of {\WF} and the coherent fraction depend on the temperature and the molecule's partial pressure. Panels (c) and (d) show the temperature dependence (at a constant pressure of $10^{-6}~Pa$) of both properties separated into contributions of {\onoffmixedstate} {\structures}. Regarding Figure \ref{fig:phase_diagram}c, we note that initially one would expect that the {\WF} of {\mixedstate} {\structures} lies between that of {\onoffstate} {\structures}. However, {\mixedstate} {\structures} have the lowest expectation values for {\WF} below $200~K$ and the highest one above. This results from the fact that here the adsorption energies do not correlate with {\WF} and {\onstate} {\structures} with large {\WF} are not energetically favorable (see Supporting Information).

Within the phase diagram we define a number of experimentally accessible temperatures and pressures at which we readout the interface state. Pressure in our case refers to the partial pressure of the TCP molecules in gas phase. Since this is difficult to control in practice, we will primarily switch our interface using temperature. For the discussion we consider temperatures of common coolants, namely that of liquid helium ($\approx 4~K$), that of liquid nitrogen ($\approx 77~K$) and room temperature ($\approx 300~K$).

At $4~K$, the thermal occupation is dominated by higher-order commensurate {\offstate} {\structures} ({\offstate} interface). Here we predict a {\WF} of approximately $-600~meV$, which matches the contribution from  {\offstate} {\structures} in Figure \ref{fig:phase_diagram}c. Since molecules in {\offstate} {\structures} adsorb at similar adsorption heights and are undistorted, the expectation value for the coherent fraction is $1.00$ (see Figure \ref{fig:phase_diagram}d).

At $77~K$, {\mixedstate} {\structures} dominate the thermal occupation ({\mixedstate} interface). Here we find a {\WF} of approximately $-650~meV$ in accordance with the expectation value for {\mixedstate} {\structures} in Figure \ref{fig:phase_diagram}c. The fact that these {\structures} contain both {\onoffstate} {\geoms} (which adsorb at different heights) leads to a decrease of the coherent fraction to about $0.67$. 

At $300~K$, a large number of {\structures}, including {\offmixedstate} {\structures}, contribute to the thermal occupation (see Supporting Information). However, {\onstate} {\structures} constitute the majority ({\onstate} interface). The {\WF} lies between $-550~meV$ and $-500~meV$. The coherent fraction at $300~K$ is approximately $0.70$. Furthermore, at partial pressures lower than approximately $10^{-7}~Pa$ it is no longer energetically favorable for molecules to adsorb on the substrate at all (indicated by the hatched areas in Figure \ref{fig:phase_diagram}).

In passing, we must discuss the uncertainty of our prediction method and the impact this has on the phase diagram. The uncertainty of our adsorption energy predictions is approximately $0.04~eV$ per molecule. The small energy differences between {\offmixedstate} {\structures} lie within this uncertainty. This mostly affects the {\offstate} interface, where only very few {\structures} contribute to the thermal occupation. At higher temperatures a larger number of {\structures} contribute to the thermal occupation averaging out the prediction error. A detailed investigation of the (minor) influence of those errors on our phase diagrams is provided in the Supporting Information.

Based on our results, two ways of switching the interface exist: A switch from the {\offstate} to the {\mixedstate} interface shifts the coherent fraction by approximately $0.3$. Switching from the {\mixedstate} to the {\onstate} interface shifts {\WF} by more than $100~meV$. This is in the order of magnitude of the typical threshold voltage of a germanium diode ($200~mV$), demonstrating the possibility of using TCP on Pt(111) to construct a switchable interface.

\begin{figure}[tbph!]
	\includegraphics[width=1\linewidth]{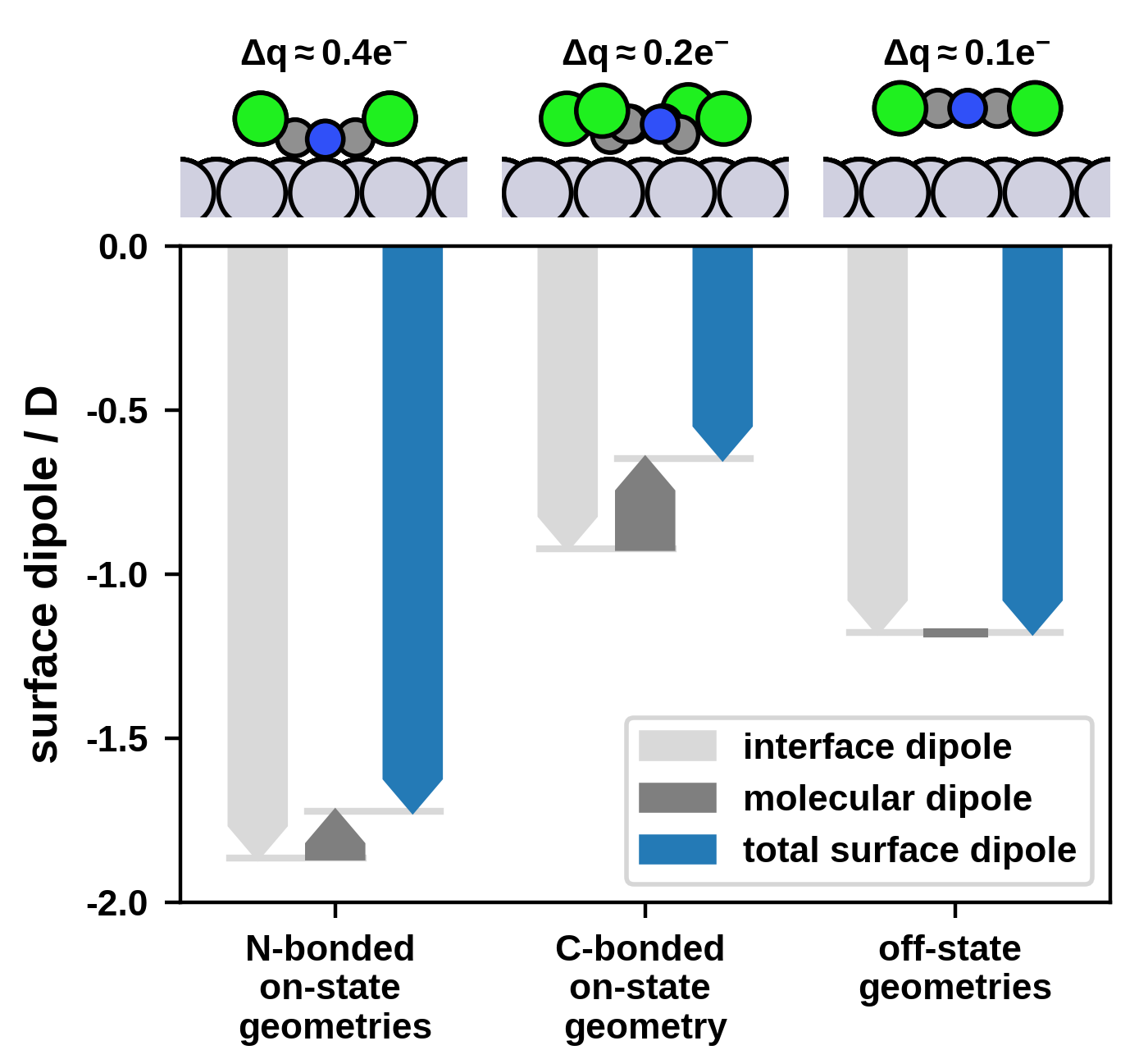}
	\caption{Surface dipoles and (Mulliken) change transfer for {\onoffstate} {\geoms}.}
	\label{fig:surface_dipole}
\end{figure}

However, the amount by which we can switch {\WF} is an order of magnitude smaller than the range of possible {\WFs}. This is partly due to the thermal occupation of energetically higher-lying {\structures} leading to an averaged interface property. However, the main reason comes to light when looking closely at the adsorption geometries and their surface dipoles. {\Onoffstate} {\geoms} exhibit significantly different adsorption geometries leading to different surface dipoles (see Figure \ref{fig:surface_dipole}). 

{\Offstate} {\geoms} are very similar and have a surface dipole of $-1.2~D$ (data shown for most stable  {\offstate} {\geom}). Since they are physisorbed, interface charge transfer is small with molecules receiving a Mulliken charge of $0.1$ electrons.

Conversely, {\onstate} {\geoms} can be separated into two significantly different groups: Three {\onstate} {\geoms} bind to the surface via N-atoms (N-bonded) while the fourth binds via C-atoms. The three N-bonded {\geoms} exhibit a surface dipole of approximately $-1.7~D$ and receive a Mulliken charge of $0.4$ electrons (data shown for most stable N-bonded {\geom}). The C-bonded {\geom} has a surface dipole of only $-0.6~D$ while getting a Mulliken charge of $0.2$ electrons. However, all {\onstate} {\geoms} have similar adsorption energies.

Therefore, {\onmixedstate} {\structures} containing the C-bonded {\geom} yield a small {\WF} (comparable to the {\offstate}). Since such {\structures} are energetically favorable, they contribute to the thermal occupation and decrease the magnitude of the switch in {\WF}. A possible solution would therefore be using a bi-stable system where {\onoffstate} {\geom} have similar surface dipoles within the respective state but differ when compared to each other. This could be fulfilled by a molecule that exhibits only one type of bonding chemistry which would lead to more similar {\onstate} {\geoms}.

\section{Conclusion}

In this contribution we discuss a switchable interface using the different adsorption states of TCP on Pt(111). We perform surface structure search and find three different classes of {\structures} that exhibit different {\WFs} and coherent fractions. Using temperature and pressure as handles we can switch between three interface states. A change from the {\offstate} to the {\mixedstate} interface shifts the coherent fraction by approximately $0.3$. More interestingly, switching from the {\mixedstate} to the {\onstate} interface shifts {\WF} by more than $100~meV$. This is in the order of magnitude of the typical threshold voltage of a germanium diode ($200~mV$), demonstrating the possibility of using TCP on Pt(111) to construct a switchable interface.

However, the achieved switch is small compared to the range of possible {\WFs}. Aside form the thermal occupation averaging interface properties, this has two reasons: (i) {\onstate} {\geoms} exhibit different bonding chemistry and surface dipoles and (ii) {\structures} with extreme {\WFs} are energetically unfavorable. This could be overcome by using a molecule which allows for only one type of bonding chemistry.

\section{Methods}
\label{sc:Methods}


To perform first principles calculations, we use the FHI-aims code\cite{aims} with numerical atom centered orbitals, the PBE exchange correlation functional\cite{pbe} and the TS\textsuperscript{surf} vdW correction\cite{TS, TSsurf}. We use periodic boundary conditions to model continuous layers. Since we are dealing with surfaces, we employ the repeated slab approach and decouple the unit cells vertically by using a vacuum of approximately $90~\angstrom$ as well as a dipole correction.\cite{neugebauer1992adsorbate} Furthermore, we use k-grids equivalent to the $(48, 48, 1)$ grid of the primitive substrate unit-cell.

The property we primarily consider is the adsorption energy (or bonding energy) $E_{ads}$. We define $E_{ads}$, using the total energy of the combined system $E_{mol+sub}$, the energy of a tetracholropyrazine molecule in vacuum $E_{mol}$ and the energy of the clean Pt(111) substrate $E_{sub}$.
\begin{equation}
	E_{ads} = E_{mol+sub} - E_{mol} - E_{sub}
\end{equation}


To perform structure search for commensurate interfaces we use the SAMPLE approach.\cite{sample} SAMPLE uses coarse-graining and Bayesian linear regression. The key premise of this approach is that the unit cell of the adsorbate layer is a super cell of the substrate. This allows us to generate all possible {\structures} within a given range of coverages and a limited number of molecules per unit cell. The first step of building {\structures} is generating an exhaustive list of substrate super cells. Then we place molecules into these unit cells. Hereby we coarse-grain the possible positions and orientations of the molecule on the surface. Therefore, we determine the local minima geometries of the isolated molecule on the substrate. This yields a small number of molecular geometries that we can place in the substrate super cells to assemble {\structures}. The number of possible {\structures} we regularly deal with is in the order of $10^6$. Since first principles calculations cannot be done on all these {\structures} we use an energy model (equation (\ref{eq:energy_model})) to determine the adsorption energies $E_{ads}$. The energy model consists of molecule-substrate and molecule-molecule interactions ($U_i$ and $V_p$). The sum over all interactions that occur in a {\structure} then yields its energy $E_{ads}$.

\begin{equation}
	E_{ads} = \sum_i n_i U_i + \sum_p n_p V_p
	\label{eq:energy_model}
\end{equation}

The interactions $U_i$ and $V_p$ are initially unknown. To determine them, we use a type of machine learning called Bayesian linear regression. Hereby, we compute a small, D-optimally selected, number of {\structures} with DFT. Using these calculations, the Bayesian linear regression algorithm then learns the molecule-substrate and molecule-molecule interactions, which allows predicting the energies of all {\structures}.


Since we expect that tetrachloropyrazine on Pt(111) forms incommensurate layers we need the ability to also predict such {\structures}. Therefore, we generalize the SAMPLE approach ({\SAMPLEGPR}). We replace the energy model with a Gaussian process, which drops the requirement of discrete molecule-substrate and molecule-molecule interactions. This algorithm is similar to that described in a previous publication.\cite{hormann2020reproducibility} GPR based algorithms have been used before to find adsorption geometries of individual molecules on substrates.\cite{todorovic2019bayesian} Our algorithm can handle isolated molecule on the surface as well as continuous layers. Put simply, a GPR algorithm is a sophisticated method to interpolate adsorption energies and work functions (or other scalar properties). Hereby the key assumption is that two {\geoms}/{\structures} that are geometrically similar have similar properties. The trick is finding a good measure for this similarity. We use radial distance functions (RDF) $\boldsymbol{f}$ which contain interatomic distances between molecules and the substrate as well as interatomic distances between molecules. To determine the similarity $C_{\alpha\beta}$ of two {\geoms}/{\structures}, $\alpha$ and $\beta$, we only need to calculate the overlap integral between the two RDFs $\boldsymbol{f_{\alpha}}$ and $\boldsymbol{f_{\beta}}$.

\begin{equation}
C_{\alpha\beta} = \langle \boldsymbol{f_{\alpha}} , \boldsymbol{f_{\alpha}} \rangle = \int \boldsymbol{f_{\alpha}}(x) \cdot \boldsymbol{f_{\beta}^*}(x) ~\mathrm{d}x
\end{equation}

The RDFs are normed such that $\langle \boldsymbol{f_{\alpha}} , \boldsymbol{f_{\alpha}} \rangle = 1$. Therefore, $C_{\alpha\beta}$ is $1$ for identical {\geoms}/{\structures} and yields smaller values for dissimilar {\geoms}/{\structures}. 

Like SAMPLE, {\SAMPLEGPR} also requires training data. For isolated molecules we evaluate the model uncertainty and choose the data points with the largest uncertainty. In case of continuous adlayers we reuse the training sets from the SAMPLE approach.

To optimize {\structures} we employ simulated annealing. Hereby {\SAMPLEGPR} provides the energy predictions. Our algorithm can optimize the most relevant degrees of freedom, namely the unit cell parameters, as well as the positions and orientations of all molecules in the unit cell.

So far we have only discussed surface structure search relying on energies from DFT calculations. These energies do not account for the effects of temperature and pressure, which are vital for a comparison with experiment. To model the impact of different thermodynamic conditions, we use ab-initio thermodynamics.\cite{rogal_reuter} Hereby we consider the thermodynamic equilibrium at a given temperature and pressure making the Gibbs free energy of adsorption the measure of interest. When determining the Gibbs free energy we neglect the contributions of the vibration enthalpy, the configuration entropy and the mechanical work as is commonly done in literature.\cite{rogal_reuter, herrmann2015structure, reuter2001composition} Using the Gibbs free energy, we can compute the probability for each {\structure} to occur at a given temperature and pressure. All probabilities combined yield the thermal occupation, which we can use to determine expectation values for {\WFs} and coherent fractions. Therefore, we calculate the mean weighted by the thermal occupation.

For additional details regarding the methods we refer to the Supporting Information.

\begin{acknowledgement}

Financial support by the FWF START award (Y 1157-N36) is gratefully acknowledged. Computational results have been achieved using the Vienna Scientific Cluster (VSC).

\end{acknowledgement}

\begin{suppinfo}

The Supporting Information contains convergence tests, additional explanations of the methods used as well as additional details regarding intermediate and final results including phase diagrams and uncertainty estimations.

\end{suppinfo}

\section{Data Availability}

The data that support the findings of this study are openly available in the NOMAD repository at \href{https://nomad-lab.eu/prod/rae/gui/dataset/id/WQfZt1jPQqC8gSIOU-tocA}{https://nomad-lab.eu/prod/rae/gui/\\dataset/id/WQfZt1jPQqC8gSIOU-tocA}.

\bibliography{bibliography}

\end{document}